\documentclass[aps,prl,twocolumn,showpacs,superscriptaddress]{revtex4-1}
\usepackage{amsmath,amssymb}
\usepackage{graphics,graphicx,color}
\usepackage{dcolumn,bm}
\usepackage{tikz}

\def\[{\left[}
\def\]{\right]}
\def\({\left(}
\def\){\right)}
\def\be{\begin{equation}}
\def\ee{\end{equation}}
\def\bea{\begin{eqnarray}}
\def\eea{\end{eqnarray}}
\def\Q{\mathbf{Q}}
\def\S{\mathbf{\sigma}}
\def\v{\mathbf{v}}
\def\r{\mathbf{r}}
\def\O{\mathbf{\Omega}}
\def\U{\mathbf{U}}

\newcommand{\iisc}
{\affiliation{Centre for Condensed Matter Theory, Department of Physics, Indian Institute of Science, Bangalore 560012, India}}

\newcommand{\imsc}
{\affiliation{The Institute of Mathematical Sciences, Chennai 600113, India}}

\newcommand{\ncbs}
{\affiliation{Simons Centre for the Study of Living Machines, National Centre for Biological Sciences (TIFR), Bangalore 560065, India}}

\begin{document}
\title
{Glassy swirls of active dumbbells}

\author{Rituparno Mandal}%
\email[Email: ]{rituparno@physics.iisc.ernet.in}
\iisc

\author{Pranab Jyoti Bhuyan}%
\email[Email: ]{pranab@physics.iisc.ernet.in}
\iisc

\author{Pinaki Chaudhuri}%
\email[Email: ]{pinakic@imsc.res.in}
\imsc

\author{Madan Rao}%
\email[Email: ]{madan@ncbs.res.in}
\ncbs

\author{Chandan Dasgupta}%
\email[Email: ]{cdgupta@physics.iisc.ernet.in}
\iisc

\begin{abstract}
The dynamics of a dense binary mixture of soft dumbbells, each subject to an active propulsion
force and thermal fluctuations, shows a sudden arrest, first to a translational then to
a rotational glass, as one reduces temperature $T$ or the self-propulsion force $f$.
Is the temperature-induced glass different from the activity-induced glass ? To address this question, we
monitor the dynamics along an iso-relaxation-time contour in the $(T-f)$ plane. 
We find dramatic differences both in the fragility and in the nature of dynamical heterogeneity which characterise
the onset of glass formation - the activity-induced glass exhibits large swirls or vortices, whose
scale is set by activity, and appears to diverge as one approaches the glass transition. 
This large collective swirling movement should have implications for collective cell migration in epithelial layers.
\end{abstract}

\maketitle

Assemblies of self-propelled objects jam at high
densities~\cite{kurchan:13, berthier:14, ni:13, marchetti:11} and
low temperatures~\cite{mandal:16}.  On approaching dynamical arrest
from the fluid side, these dense active assemblies are seen to
exhibit typical glassy dynamics, with activity manifesting simply
as an {\it effective temperature}~\cite{mandal:16, flenner:16}.
Likewise, starting from the jammed state, activity is seen to
prematurely {\it fluidize} the system at a reduced (enhanced)
transition temperature (volume fraction)~\cite{kurchan:13, berthier:14,
ni:13, marchetti:11, mandal:16}. On the face of it, it might appear
that an active glass behaves very similar to a conventional one,
albeit with a different effective temperature or density~\cite{kurchan:13,
flenner:16}. In this paper, we provide evidence to the contrary -
we show that an active glass exhibits distinctive dynamical features
on account of their local driving.

To address this, we study  dense assemblies of generic oriented
nonspherical self-propelled objects~\cite{suma:15, tung:16,
wensink:PNAS12, wensink:12, siebert:16, hinz:15}, which are free
to explore both  translational and orientational degrees of freedom.
Living realisations include reconstituted layer of confluent
epithelial cells~\cite{manning1:15}, where cell shape anisotropy
plays a crucial role in the jamming-unjamming
transition~\cite{manning2:15}, and jammed biofilms formed by a dense
collection of rod-shaped bacteria. Likewise, shaken non-spherical
grains at high packing densities constitute non-living
examples~\cite{kumar:14}.

\begin{figure*}
\centering
\includegraphics[height = 0.28\linewidth]{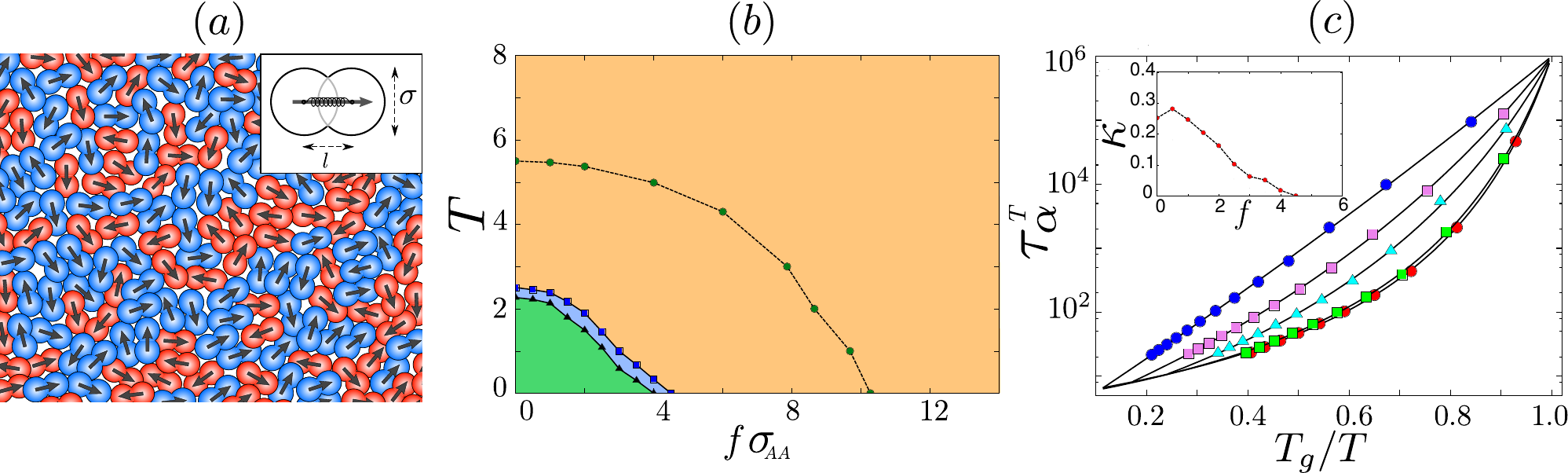}
\caption{(color online) (a) Schematic of a dense binary assembly
of self-propelled dumbbells of $A$ (blue) or $B$ (red) type. Inset
shows structural details of each dumbbell - two identical soft
Lennard-Jones spheres of diameter $\sigma$ connected through a
simple spring of spring constant $k$ and rest length $l$. The
body-fixed self-propulsion force $f$ (arrow) is along the long axis
of each dumbbell. (b) Phase diagram in ($T-f$) plane shows a liquid
(L, in brown), translational glass (TG, in blue) and
translational-rotational glass (TRG, in green) with phase boundaries
determined from VFT fits. The dashed line represents the iso-relaxation time
($\tau^{T}_{\alpha}=102.85$) line, along which the dynamics is
probed at points marked in green. (c) Variation of translational
relaxation times $\tau^{T}_{\alpha}$ versus scaled temperature
$(T/T_g)$ with changing active force : $f=0$ (red), $f=1$ (green),
$f=2$ (cyan), $f=3$ (violet), $f=4$ (blue). $T_g$ is defined as the
temperature where $\tau^{T}_{\alpha}=10^6$.  Inset shows how the
corresponding kinetic fragility $\kappa$ decreases with increased
self-propulsion force.}
\label{fig:schematic}
\end{figure*}

\begin{figure}
\centering
\includegraphics[height = 0.8\linewidth]{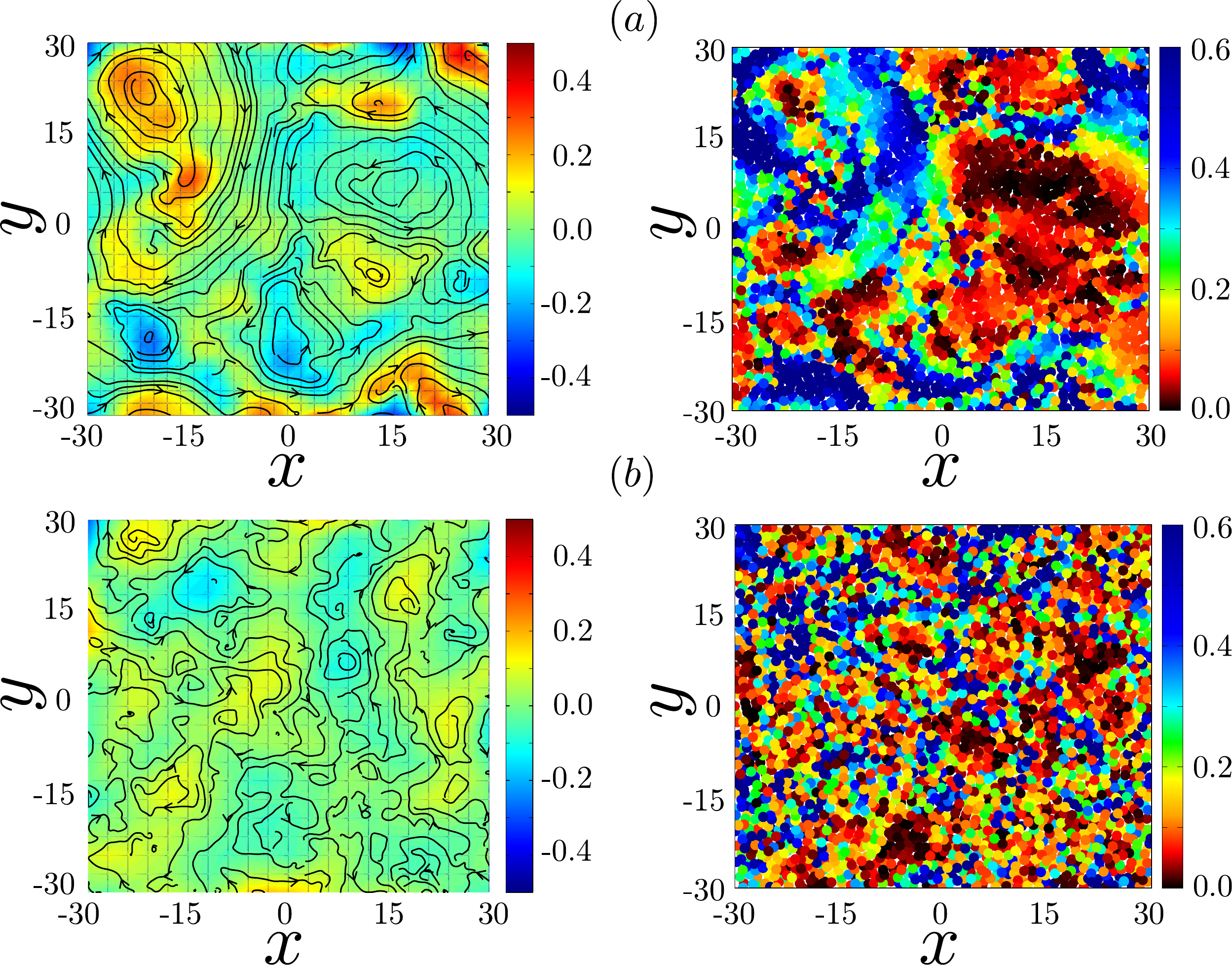}
\caption{(color online). (a) ({\it left}) Streamlines of displacement
field, ${\bf d}(\r)$, for the active dumbbells $(f=10.3, T=0)$
computed over $t= \tau^{T}_{\alpha}$ showing large vortex-like
structures with the underlying colormap reflecting the corresponding
values of the vorticity $\omega(\r)$. ({\it right}) Map showing
the magnitude of displacement during $t= \tau^{T}_{\alpha}$, illustrating the
extent of heterogeneous dynamics in the active glass, with blue
particles being the fastest and black particles being the slowest.
Note that the vorticity and particle velocities are {\it anti-correlated}
(see text). (b) ({\it left}) Streamlines of displacement for passive
dumbbells $(f=0,T=5.5)$, along with corresponding vorticity colormap.
({\it right}) Displacement map for the passive liquid. Note the
absence of any large scale structures and the resultant low vorticity.}
\label{fig:displacement}
\end{figure}

\begin{figure}
\vspace{4mm}
\includegraphics[height = 0.47\linewidth]{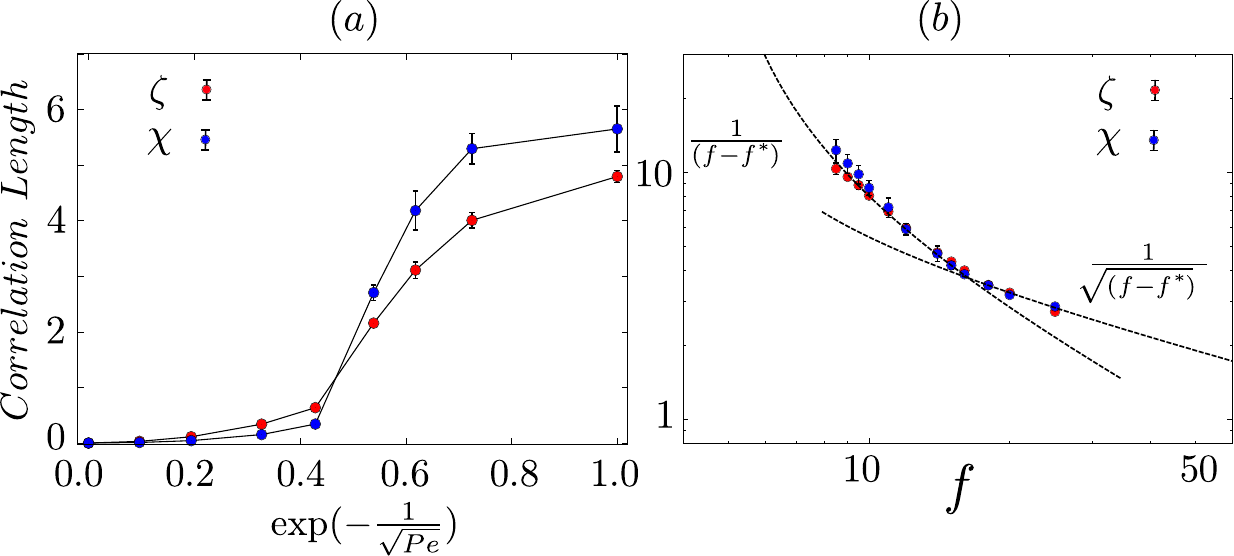}
\caption{(color online). (a) Correlation lengths $\zeta$,
characterizing the degree of alignment of the displacement vectors, and $\chi$,
characterizing the spatial corelation of vorticity, as a function of $Pe$, along
the iso-$\tau^{T}_\alpha$ line, show a crossover from a passive to
an active supercooled liquid. To highlight the change,  we subtract
from the measured correlation lengths, a microscopic length
associated with the size of the dumbbell.
The choice of plotting against $\exp[{-1/\sqrt{Pe}}]$ ensures
uniform spacing for the simulation data  points over the range of $Pe$.
 (b) $\zeta$ and $\chi$ as a function of activity $f$ for $T=0$.
 These lengths increase and show a crossover from  $1/\sqrt{f-f^*}$ dependence at
 large $f$ to $1/(f-f^*)$ dependence close to the glass transition at $f^*=4.5$.}
\label{fig:lengthscale}
\end{figure}

We perform Brownian dynamics simulations of a dense binary assembly
of  dumbbells~\cite{kob:94, chong:05, moreno:05, chong:09}
in 2-dimensions (see Fig.\,\ref{fig:schematic}(a) and {\it Supplementary
Information}~\cite{supplement} for details); the 50:50 mixture of  \textit{A} and
\textit{B} type dumbbells ensures amorphous steady state structures.
This assembly, subject to a temperature bath $T$, is made active
by driving each dumbbell with a body-fixed propulsion force $f$
along the long axis of each dumbbell, and is characterized by a
Pe\'clet number $Pe\equiv f\sigma_{AA}/k_BT$, measuring the relative
strength of activity with respect to temperature~\cite{suma:15,hinz:15}.
The phase diagram Fig.\,\ref{fig:schematic}(b) shows a jammed state
upon reducing either temperature or self-propulsion force. Our main
result is that the dynamical signatures of an active glass are
fundamentally different from a conventional glass - (i) activity
makes the glass less fragile (Fig.\,\ref{fig:schematic}(c)), and
(ii) the nature of dynamical heterogeneity in an active glass is
very unique and exhibits large scale swirls and vortices
(Fig.\,\ref{fig:displacement}(a)), whose size increases and appears
to diverge as one approaches the arrested state
(Fig.\,\ref{fig:lengthscale}(b)) by reducing the self propulsion force
$f$. We understand these length scales from a continuum hydrodynamic
theory of active dumbbells. These large scale swirls are the sluggish
imprints of the collective turbulent motion observed in an active
fluid of anisotropic particles~\cite{wensink:PNAS12, wensink:12, giomi:15}.

To identify the onset of translational and rotational glassy behaviour
in the ($T-f$) plane, we monitor  the mean-square-displacement
($MSD$), $\langle \Delta \r(t)^2 \rangle = \langle\frac{1}{N} \sum_i
\langle \vert {\bf r}_i(t_0+t) - {\bf r}_i(t_0)\vert^2 \rangle$,
where $\bf{r}_i$ is the center-of-mass of the $i^{th}$-dumbbell;
the orientation correlation function $C_2(t) = \frac{1}{N}
\sum_{i}\langle P_2\left[{{\bf n}}_i(t_0)\cdot{{\bf
n}}_i(t_0+t)\right]\rangle$~\cite{chong:05, moreno:05}, where $P_2$
is the $2^{nd}$-order Legendre polynomial and ${\bf n}_i$ is the
unit vector along the long axis of the $i^{th}$ dumbbell; and the
overlap function $Q(t) = \langle\frac{1}{N} \sum_{i} w(\vert{\bf
r}_i(t_0)- {\bf r}_i(t_0+t)\vert)\rangle$, where $w(r)=1$  if  r
$\leq$ a  and 0 if r $>$ a, with a=0.3 \cite{smarajit:09}. Here,
$\langle\cdots\rangle$ denotes an average over time origins $t_0$
and trajectories and $N$ is the number of dumbbells.

Note the following - (i) the dumbbells are subject to thermal noise
and active propulsion; (ii) there is no explicit activity decorrelation
time, orientational decorrelation occurs because of collisions and
thermal fluctuations; and (iii) there is no imposed orientational
alignment, any kind of orientation correlation appears
entirely due to packing and collisions.

Starting from the liquid phase at high temperature $T$ and activity
$f$, we reduce either $T$ or $f$. The MSD ({\it Supplementary Figure
S1} \cite{supplement}) shows cage diffusion, with a distinct intermediate plateau and
a reduced late time diffusion coefficient, characteristic of the
approach to a glass, as the temperature or activity is reduced. The
boundary to the glassy phase is associated with the vanishing of
the late time diffusion coefficient. This is accompanied by a slowing
down of both the translational and rotational structural relaxation
process, captured by the two-step decay of the   correlation
functions, $Q(t)$ and $C_2(t)$ ({\it Supplementary Figure S2}
\cite{supplement}).   Thus, the corresponding $\alpha$-relaxation
times $\tau^{T}_{\alpha}$ and $\tau^{R}_{\alpha}$ increase, as $T$
or $f$ decreases ({\it Supplementary Figure S3} \cite{supplement}).
As is usual in studies of glass forming systems, we estimate the
glass transition temperature, separately for both rotational and
translational relaxation, by fitting the respective $\alpha$-relaxation time
to a Vogel-Fulcher-Tammann (VFT) form, 
$\tau_{\alpha}=\tau_\infty \exp \[\frac{1}{\kappa\(\frac{T}{T_{\mbox{\tiny{VFT}}}}-1\)} \]$, 
where $\tau_\infty$ is the relaxation time at large temperatures,
$\kappa$ is the coefficient of kinetic fragility, and
$T_{\mbox{\tiny{VFT}}}(f)$ is the putative glass transition temperature
for an applied active force $f$. The resultant phase diagram in the
$T-f$ plane, obtained by marking the different $T_{\mbox{\tiny{VFT}}}(f)$
values, is shown in Fig.\,\ref{fig:schematic}(b). We find that the
rotational and translational degrees of freedom freeze out from the
liquid at different $T_{\mbox{\tiny{VFT}}}(f)$, resulting in
two distinct glass phases, the translational glass (TG) and
translational-rotational glass (TRG).

The phase diagram itself does not reveal any difference between
approaching the glass by lowering $T$ or $f$. To see differences
in the active and passive systems, distinguished by their Pe\'clet
number, one needs to probe their dynamical heterogeneity.  Already,
the VFT fits give some hint of this: the kinetic fragility ($\kappa$)
of translational glass shows a decrease with increasing activity
$f$; see Fig.\,\ref{fig:schematic}(c) and inset.  A similar behaviour
is also observed for the rotational relaxation (see {\it Supplementary
Figure S4} \cite{supplement}).  Thus, the active glass becomes
{\textit{stronger}} or less {\textit{fragile}} than a passive glass,
both during  translational and rotational arrest.  This allows for
tuning of the glassy behaviour from being fragile to strong, by
tuning the magnitude of active propulsion. The decrease in fragility
in an active soft supercooled liquid is consistent with some earlier
numerical~\cite{mandal:16} and theoretical~\cite{nandi:16} studies.  We will see later
how this result is  also consistent with observations in reconstituted
epithelial tissues that have been reported to show glassy
behaviour~\cite{zhou:09}.

To explore the difference in behaviour while approaching the arrested
state by either decreasing $T$ or $f$, we probe the dynamics in
finer detail at specific points along an iso-relaxation time
($\tau^{T}_{\alpha}=102.85$) line (see Fig.\,\ref{fig:schematic}(b)),
which lies in the supercooled liquid regime. We measure the
displacement field vectors of the centre of mass of the dumbbells,
${\bf d}(\r)$, over a time $\tau_{\alpha}^T$, and construct a spatial
map of  corresponding streamlines, which are shown in the left panel
of Fig.\,\ref{fig:displacement}. For the passive system ($f=0$),
such a map is structureless with no large scale spatial correlation, Fig.\,\ref{fig:displacement}(b)(left panel).  On increasing $Pe$ along
the iso-$\tau^{T}_{\alpha}$ line, one suddenly begins to observe
distinct \textit{vortex} like structures, whose size increases with
increasing $Pe$, as seen in Fig.\,\ref{fig:displacement}(a)(left panel).
This pattern (or lack of it in the passive case) is also quantified
via the coarse-grained vorticity $\omega(\r)=\nabla \times {\bf
v}_{\tau}(\r)$, where ${\bf v}_\tau(\r) \equiv {\bf d}(\r)/\tau_{\alpha}^T$~\cite{note}, which are shown as underlying colormaps in the left panels
of Fig.\,\ref{fig:displacement}. Further, in the right panel of
Fig.\,\ref{fig:displacement}, we show spatial maps of the magnitude
of single particle displacements during $\tau_{\alpha}^T$, to gauge
the extent of dynamical heterogeneity exhibited in either situation.
The active system, Fig.\,\ref{fig:displacement}(a)(right panel), has
much more spatial heterogeneity in dynamics than the passive one, Fig.\,\ref{fig:displacement}(b)(right panel). Also, notice the
anti-correlation in the vorticity and the magnitude of particle
displacement: regions with low (high) vorticity correspond to fast
(slow) dumbbells, particles between a vortex-antivortex pair are
faster, while particles in the vortex core are slower.

In order to extract correlation lengths from the spatial structures
visible in such maps, we calculate the angle-averaged correlation
functions of  (i) the orientation of the displacement vectors of
the dumbbell,  $C(r)=\langle 2 \cos^2\Delta \theta(\r)-1\rangle$,
where $\Delta \theta(\r)$ is the angular separation between two
displacement vectors separated by distance $\r$ ({\it Supplementary
Figure S5} \cite{supplement}), and (ii) the vorticity, $G(r)=\langle
\omega(\mathbf{0})\omega(\r)\rangle$, evaluated over $\tau^T_\alpha$
({\it Supplementary Figure S6} \cite{supplement}). The extracted
correlation lengths, $\zeta$ and $\chi$, respectively , show a
crossover as we move along the iso-$\tau^{T}_\alpha$ line, and
distinguish the passive (low $Pe$) from the active (high $Pe$)
supercooled liquids; see Fig.\,\ref{fig:lengthscale}(a).

If we now move towards the dynamically arrested regime from either
extreme ends, i.e.  along the passive direction  ($Pe=0$) where the
active forcing is absent or the athermal active  direction ($Pe=\infty$)
where thermal fluctuations are suppressed, we clearly observe the
stark differences in the way the above-mentioned spatial correlations
grow.  As one goes towards the glass transition at $Pe=0$,  there
is no significant change in the correlation lengths $\zeta$ and
$\chi$ -  which continue to remain at the scale of the dumbbell -
({\it Supplementary Figure S7} \cite{supplement}). There is of
course the usual dynamical heterogeneity  associated with the
emergence and growth of {\textit{fast}} moving and {\textit{slow}}
moving domains \cite{berthier:11} and corresponding  increase in
relaxation times \cite{smarajit:09, smarajit:14}.  However, on
approaching the glass transition from the active side, along
($Pe=\infty$), the nature of the dynamical heterogeneity is very
different and is associated with swirling or vortex
patterns that grow in size. The corresponding growing length scales (see
Fig.\,\ref{fig:lengthscale}(b)), goes as $1/\sqrt{f-f^*}$, far
away from the glassy regime ($f^*=4.5$), and crosses over to
$1/(f-f^*)$, as one nears dynamical arrest.

To understand the origin of these vorticity scales, we construct a
hydrodynamic description of the self-propelled dumbbells in the
isotropic phase.  Our simulations show that the velocity $\v_\tau$
is correlated with the orientation of dumbbell ${\bf n}$, or more
precisely, $\Q\cdot{\bf f}$, where $\Q$ is the nematic orientation
tensor describing the apolar orientation of the dumbbell and ${\bf
f}$ is the body-attached propulsion force.  This correlation increases with Pe\'clet
number along the iso-$\tau^{T}_\alpha$-line before saturating due
to packing considerations ({\it Supplementary Figure
S8}~\cite{supplement}). This suggests that the appropriate
hydrodynamic fields are the conserved densities, orientation tensor
$\Q$ and the dumbbell velocity $\v$. Note that the polarity of the
dumbbells is an acquired property, a consequence of $\Q \cdot \v$.
To see the generation of large swirls or vortices, it suffices to
consider the dynamics of  $\Q$ and ${\bf v}$ alone ({\it Supplementary
Information} \cite{supplement}). Newtons second law provides the
equation for the  dumbbell velocity $\v$,
\be
\rho \partial_t \v =  - (\Gamma - \eta\nabla^2) \v +   \nabla \cdot \S
\ee
Our use of Brownian dynamics implies that there is a frictional
damping $\Gamma$, in addition to the usual viscous damping associated
with the transfer of momentum arising from collisions. The last
term corresponds to local forces due to a stress $\sigma = \sigma^{op}
+ \sigma^{act}$, written as a sum of the order parameter stress,
$\S^{op}= \Lambda \left({\delta F}/{\delta \Q}\right) - \left[\Q \cdot ({\delta F}/{\delta \Q}) - ({\delta F}/{\delta \Q}) \cdot \Q\right]$~\cite{giomi:15},
 derived from a free-energy functional, $F[\Q]$ for the nematic
 ({\it Supplementary Information} \cite{supplement})
\be F[\Q] = \int d\vec{x}{\left[ \frac{A}{2} \Q^2+\frac{C}{4}
\Q^4+\frac{K}{2} (\nabla \Q)^2 \right]} \ee 
and $\S^{act}= f l \Q$ is the active stress \cite{marchetti:14}. 
The dynamics of
$\Q$ is given by \be D_t \Q = \lambda {\U} + \Q \cdot \O - \O \cdot
\Q - \gamma^{-1}\frac{\delta F}{\delta \Q} \ee where $D_t$ represents the
convective derivative in the comoving frame, $\lambda$ is the
stable-flow alignment parameter, $\gamma$ is the rotational viscosity and $\U$ and $\O$ are the symmetric and
anti-symmetric strain-rates. Now linearising about the isotropic state with no
flow, we obtain,
\bea 
{\dot{\Q}} & = & \frac{\lambda}{2}\left[(\nabla \v)+(\nabla \v)^T\right]-\frac{1}{\gamma}\left[A-K \nabla^2\right]  {\Q}
\label{eq:hydroq} \\ 
{\dot{\v}} & = & -(\Gamma-\eta \nabla^2)\v + \left[\Lambda(A - K \nabla^2) + f\,l \right] \nabla \cdot  {\Q}
\label{eq:hydrov} 
\eea
Taking the curl of the above equations gives us the equation for
the vorticity $\omega$. Away from the glass transition, patterning
is generated by balancing the active stress against the orientational
elastic stress; this  gives a length scale $\sqrt{\Lambda K/f\,l}$, consistent
with the scaling observed in the simulations at large
$f$ (Fig.\,\ref{fig:lengthscale} and~\cite{fielding:16}).  On
approaching the glass transition $f=f^*$, the translational motion
gets sluggish and finally arrested; it would seem that we can no
longer use an active fluid description. We argue however that when
$f > f^*$ we should still be able to use these equations, as long
as we renormalize time by $t \to t/\tau^{T}_{\alpha}$. In this highly
viscous fluid regime, the patterns are generated by balancing the
active stress with the viscous stress; this gives rise to a length
scale that goes as $1/(f-f^*)$ ({\it Supplementary
Information} \cite{supplement}).
The crossover between these two
different forms of the characteristic length with active force,
occurs at propulsion force $\propto \eta^2l/\Lambda K$.

To summarise, we study a model Brownian assembly of self-propelling
soft anisotropic particles, wherein there is neither any explicit
activity de-correlation time nor imposed orientational alignment,
unlike many other model active systems that have been studied. We
demonstrate that the active forcing makes the corresponding supercooled
fluid less fragile and more importantly the heterogeneous dynamics
observed in the active fluid is very different from its passive counterpart, exhibiting large-scale
swirls and vortices. We further show that the correlation lengths,
associated with these spatial structures, grow with decreasing
active forcing and appear to diverge as the dynamics gets arrested.
We rationalise the occurrence of such length scales using a continuum
hydrodynamic theory of active dumbbells.

Our study of dynamical heterogeneities in active glass composed of
anisotropic particles has implications for the collective cell
dynamics in epithelial layers. This can be seen, by  constructing
a Voronoi tessellation around each dumbbell; the resulting pattern
then resembles an epithelial sheet ({\it Supplementary Movie} \cite{supplement}).
One of the first studies of glassy dynamics close to the
jamming transition in a reconstituted epithelial sheet~\cite{zhou:09},
revealed two striking features : (i) the epithelial glass was
significantly less fragile  compared to a conventional hard-sphere
glass and (ii) systematically reducing cell motility by treatment
with titrated amounts of actin de-polymerization agents rendered
the epithelial sheet more fragile.  Our result on how active
propulsion makes the jammed state stronger or less fragile,
is entirely consistent with this.  In addition, our work  suggests the possibility of
observing large scale swirls in situations where cellular propulsion
is high and the cell-substrate adhesion is weak ({\it
Supplementary Movie} \cite{supplement})~\cite{Angelini:11}.

The swirling patterns or vortex structures that we observe for dense
assemblies of anisotropic particles, are reminiscent of the dynamical
patterns seen in {\it active turbulence} exhibited by  a collection
of fluid of active rods~\cite{wensink:PNAS12,wensink:12,giomi:15}.
Is the dynamical heterogeneity exhibited by the active glass a
frozen memory of its active turbulent past?  Our work shows
unambiguously that an active glass is fundamentally distinct and not
just a conventional glass with another name.

We thank A. Das, J. P. Banerjee, S. S. Ray and S. Sastry
for useful discussions. RM and PJB acknowledge financial
support from CSIR, India, and CD  from DST, India.


\begin{thebibliography}{99}
\bibitem{note} This differs from the usual definition based on
instantaneous velocities  and reflects the fact that dynamics in
the supercooled regime manifests only over $\alpha$-relaxation
times.
\end{thebibliography}
\end{document}


\title
{Glassy swirls of active dumbbells - Supplementary Information}

\author{Rituparno Mandal}%
\email[Email: ]{rituparno@physics.iisc.ernet.in}
\iisc

\author{Pranab Jyoti Bhuyan}%
\email[Email: ]{pranab@physics.iisc.ernet.in}
\iisc

\author{Pinaki Chaudhuri}%
\email[Email: ]{pinakic@imsc.res.in}
\imsc

\author{Madan Rao}%
\email[Email: ]{madan@ncbs.res.in}
\ncbs

\author{Chandan Dasgupta}%
\email[Email: ]{cdgupta@physics.iisc.ernet.in}
\iisc

\maketitle

\section{Model and Simulation Details}
For our study, we perform  Brownian dynamics simulation of a
two-dimensional binary dumbbell mixture.  Each dumbbell consists
of two spherical monomers of the same type (either A or B) connected
via a spring with a stiffness of $k=200$; a schematic of the dumbbell
is illustrated in the inset of {\textit{Fig.1}} in the main paper. This value of
spring constant makes the dumbbells fairly rigid with an equilibrium
length $l=\sigma_{\alpha \alpha}/2$ (where $\alpha\in A, B$)
between the centers of the two monomers of each dumbbell \cite{chong:05, hinz:15,
suma:15}. The mixture consists of 50:50 \textit{A-A} and \textit{B-B}
dumbbells, which leads to amorphous structures. The equation of
motion for each monomer can be written as,

\begin{equation}
{\dot{\mathbf{r}}}_i=\frac{1}{\gamma} \left[ \sum_{j\in \mathbb{N} \mathbb{N} } \mathbf{f}_{ij} -k(\mathbf{r}_i-\mathbf{r}_{i^\prime}-l\mathbf{n}_i)+ f \mathbf{n}_i  \right]+{\boldsymbol{\eta}}_i 
\end{equation}
where $\gamma$ is the friction coefficient, $\mathbf{f}_{ij}$ is the
interaction force between the $i$-th and the $j$-th monomer, $i$
and $i^\prime$ denote the two monomers of the same dumbbell,
$\mathbf{n}_i$ is the unit vector in a pre-specified direction along
the long axis of the dumbbell associated with the $i$-th particle,
$f$ is the strength of the propulsion force and $\boldsymbol{\eta}_i$
is the thermal noise which satisfies fluctuation-dissipation relation
of the form $\langle \eta^{\alpha}_i(t) \eta^{\beta}_i({t^{\prime}})
\rangle= 2 D \delta_{\alpha \beta} \delta(|t-t^{\prime}|)$ (where
$\alpha, \beta \in x,y$) with $\langle \boldsymbol{\eta}_i(t)\rangle=0$,
$D=\frac{k_B T}{\gamma}$; $T$ is the temperature of the heat bath.
The interaction force ($\mathbf{f}_{ij}$) between the monomers is
modelled via the Lennard-Jones pair potential,
\begin{equation}
 V_{ij}(r)=4 \epsilon_{\alpha \beta} \[\(\frac{\sigma_{\alpha \beta}}{r_{ij}}\)^{12}-\(\frac{\sigma_{\alpha \beta}}{r_{ij}}\)^{6}\],\label{eq:potential}
\end{equation} 
where $r_{ij}$ is the distance between the $i$-th and the $j$-th
particle {\textit{i.e.}} $r_{ij}=|\mathbf{r}_i-\mathbf{r}_j|$ where
$\alpha$, $\beta$ represent either $A$-type or $B$-type particles.
In our simulation we have chosen the values of $\sigma_{\alpha
\beta}$ and $\epsilon_{\alpha \beta}$ to be: $\sigma_{AB}=0.8
\sigma_{AA}$, $\sigma_{BB}=0.88 \sigma_{AA}$, $\epsilon_{AB}= 1.5
\epsilon_{AA}$, $\epsilon_{BB}=0.5 \epsilon_{AA}$. The potential
has been truncated at $r^{c}_{\alpha \beta}=2.5 \sigma_{\alpha
\beta}$ and the potential has been shifted accordingly such that
both the potential and the force are continuous at the cut-off. The
unit of length and energy in our simulation are set by  $\sigma_{AA}=1$
and $\epsilon_{AA}=1$ and the study is done for an  overall number
density of  $\rho=1.6$.
 
In our model,  \textit{activity} is introduced via a self-propulsion
force of magnitude $f$  which acts on each monomer, of each
dumbbell, in a pre-specified direction along the corresponding
dumbbell axis. Therefore, although each dumbbell consists of two
identical beads,  there exists a polar vector attached to each
dumbbell which breaks the head-to-tail symmetry, via the corresponding
direction of the self-propulsion force. The control parameter for
activity $f$ is also the measure of the extent of non-equilibrium
behaviour introduced in the system.

For such a system of particles, we carry out Brownian dynamics
simulation, using the Euler algorithm. The {\textit{temperature}}
$T$ for both passive and active system is determined by the heat
bath to which the system is coupled.

The number of integration step for the simulation is between $10^8$
and $10^9$ depending on the parameters with a time-step of integration
$dt=10^{-3}$. We have compared our Brownian dynamics results to
those of Newtonian dynamics simulations and observed that the long
time dynamics is quantitatively similar for the parameter ranges
we have studied, as expected. For each parameter set, the results
presented here have been averaged over $16$-$32$ independent
trajectories.

\section{Hydrodynamic calculation}

We now construct a hydrodynamic description of the self-propelled
dumbbells in terms of their conserved densities, orientation tensor
${\bf Q}$ and the dumbbell velocity ${\bf v}$. The fact that it is
a two-component system just goes towards forming a translational
and orientational glass - it is not relevant to the generation of
the large swirls or vortices. We will therefore treat it as a
one-component system with a single conserved density field $\rho$,
whose dynamics is given by,
\be
\partial_t \rho + \nabla \cdot (\rho {\bf v}) = 0
\ee
Newton's second law provides the equation for the  dumbbell velocity ${\bf v}$,
\be
\rho \partial_t {\bf v} =  - (\Gamma - \eta\nabla^2) {\bf v} - \zeta_c \nabla \rho 
+   \nabla \cdot \sigma.
\ee
$\Gamma$ represents frictional damping and $\eta$ is the viscosity associated
with the transfer of momentum arising from collisions. The second term on the right, may be thought of as
 a pressure term arising from density inhomogeneities; $\zeta_c$ is therefore a compressibility.
 The last term on the right is the contribution to the force coming from the total deviatoric stress $\sigma$.
 In the high density phase at the onset of glassy behaviour
one might choose to ignore spatial inhomogeneities of the density.
There are however non-trivial temporal correlations of the density,
but since we only want to extract the swirling behaviour, we ignore
this. This leads to Eq.\,(1) in the main text.
The total deviatoric stress $\sigma
= \sigma^{op} + \sigma^{act}$, is a sum
of the order parameter stress, $\S^{op}= \Lambda \left({\delta F}/{\delta \Q}\right) - \left[\Q \cdot ({\delta F}/{\delta \Q}) - ({\delta F}/{\delta \Q}) \cdot \Q\right]$~\cite{giomi:15},
derived from a free-energy functional
for the nematic,
\be
F[\Q] = \int d\vec{x}{\left[ \frac{A}{2} \Q^2+\frac{C}{4} \Q^4+\frac{K}{2} (\nabla \Q)^2 \right]},
\ee
and the active stress $\S^{act}= f l \Q$ \cite{marchetti:14}.
From this we get,
\be
{\delta F}/{\delta \Q}=\left[(A+{\frac{C}{2}}S^2) \Q-K \nabla^2 \Q \right]
\ee
where $S$ is the magnitude of the nematic order parameter present
in the system and $S^2=2 \mbox{Tr}(\Q^2)$. We also have the time
evolution equation for $\Q$,
\be
D_t \Q = \lambda {\U} + \Q \cdot \O - \O \cdot \Q - \gamma^{-1} \frac{\delta F}{\delta \Q}
\ee
where $D_t$ represents the convective derivative in the comoving
frame, $\lambda$ is the stable-flow alignment parameter, $\gamma$ is the rotational viscosity and $\U$
and $\O$ are the symmetric and anti-symmetric strain-rates;
$\U=\frac{1}{2}\left[(\nabla \v)+(\nabla \v)^T\right]$ and
$\O=\frac{1}{2}\left[(\nabla \v)-(\nabla \v)^T\right]$. 

Upon linearising about the isotropic state with no flow, we get equations Eqs.\,4,\,5 in the main text.
\\

\noindent
{\it Discussion on length scales} : The lengths scales associated with the spatial patterning of the vortex flows can be obtained using dimensional
analysis by balancing the
relevant contributions to the stress appearing in Eq.\,5 in the main text.

Away from the glass transition, patterning 
is generated by balancing the active stress ($f l \Q$) against the orientational
elastic stress ($\Lambda K \nabla^2 \Q$). This gives a length scale $\sqrt{\Lambda K/f\,l}$, consistent
with the scaling observed in our simulations at large $f$.  

On approaching the glass transition $f=f^*$, the translational motion
gets sluggish and finally arrested. As long as $f > f^*$, the system is still an active {\it fluid},
and flows over a time scale $\tau^{T}_{\alpha}$. 
We thus argue that we should still be able to use the active fluid equations, Eq.\,4, 5, as long
as we renormalize time by $t \to t/\tau^{T}_{\alpha}$. In this highly
viscous fluid regime, the patterns are generated by balancing the
active stress ($f l \Q$) with the viscous stress ($\eta \nabla {\bf v}$).
Note that the velocity scale that appears in the balance above is the 
coarse-grained velocity ${\bf v}_{\tau}$. From our simulations, the typical magnitude of 
${\bf v}_{\tau}$ is of order $l$, in units of this scaled time. Using this we get the length scale for 
the spatial patterning of the velocity to go as $\eta/(f-f^*)$, as long as $f$ is greater than, but close to, 
$f^*$.

The crossover between these two behaviours occurs at a propulsion force $\propto \eta^2\,l/\Lambda K$.

\clearpage

\section{Supplementary Figures}

\begin{figure}[h]
\centerline{\includegraphics[height = 0.35\linewidth,clip=true]{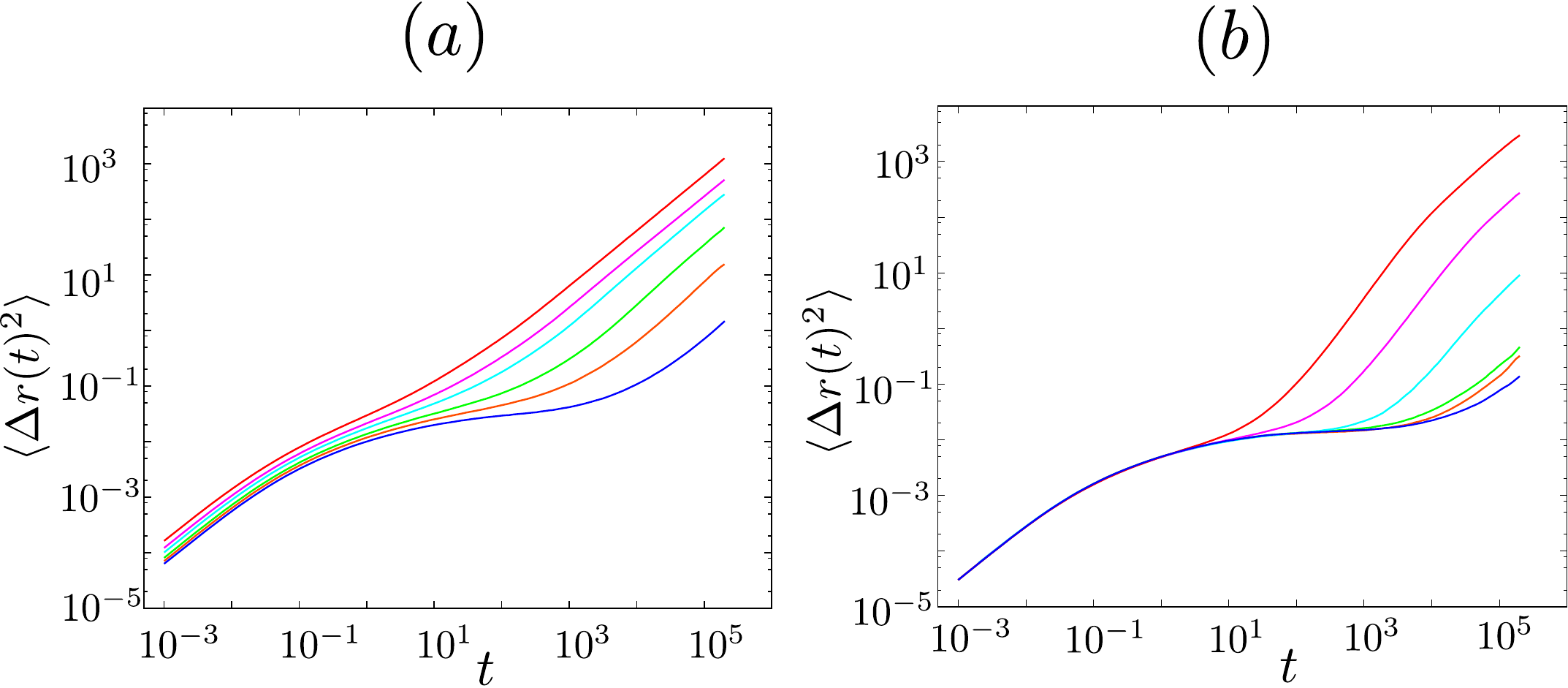}}
\caption{(color online). Mean squared displacement ($\langle \Delta r(t)^2\rangle$) as function of $t$ -  (a) for propulsion force $f$=$2.0$ for different temperatures $T$=$3$ (blue), $3.5$ (orange),$4$ (green), $5$ (cyan), $6$ (magenta), $8$ (red) and (b) for temperature $T$=$1.5$ for different propulsion forces $f$=$3$ (blue), $3.5$ (orange),$4$ (green), $5$ (cyan), $6$ (magenta), $8$ (red).}
\label{msd}
\end{figure}

\begin{figure}
\centerline{\includegraphics[height = 0.35\linewidth,clip=true]{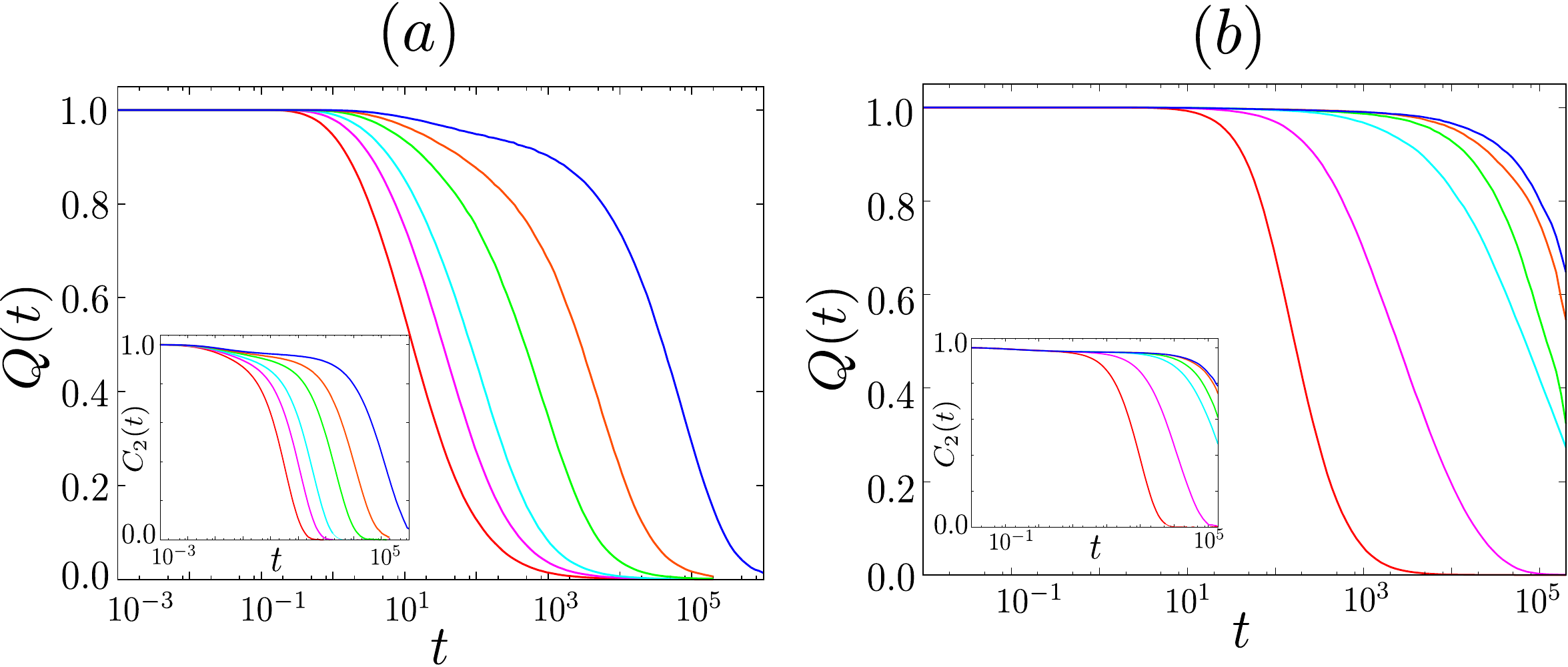}}
\caption{(color online).(a) Overlap function, $Q(t)$, measured for propulsion force $f$=$2$ at different temperatures $T$=$3$ (blue), $3.5$(orange), $4$ (green), $5$ (cyan), $6$ (magenta), $8$ (red) and (b) for temperature $T$=$1.5$ for different propulsion forces $f$=$3$ (blue), $3.5$ (orange),$4$ (green), $5$ (cyan), $6$ (magenta), $8$ (red). Inset shows rotational time correlation function, $C_2(t)$, and its behaviour for the same set of parameters in both the cases.}
\label{correlator}
\end{figure}

\begin{figure}
\centerline{\includegraphics[height = 0.4\linewidth,clip=true]{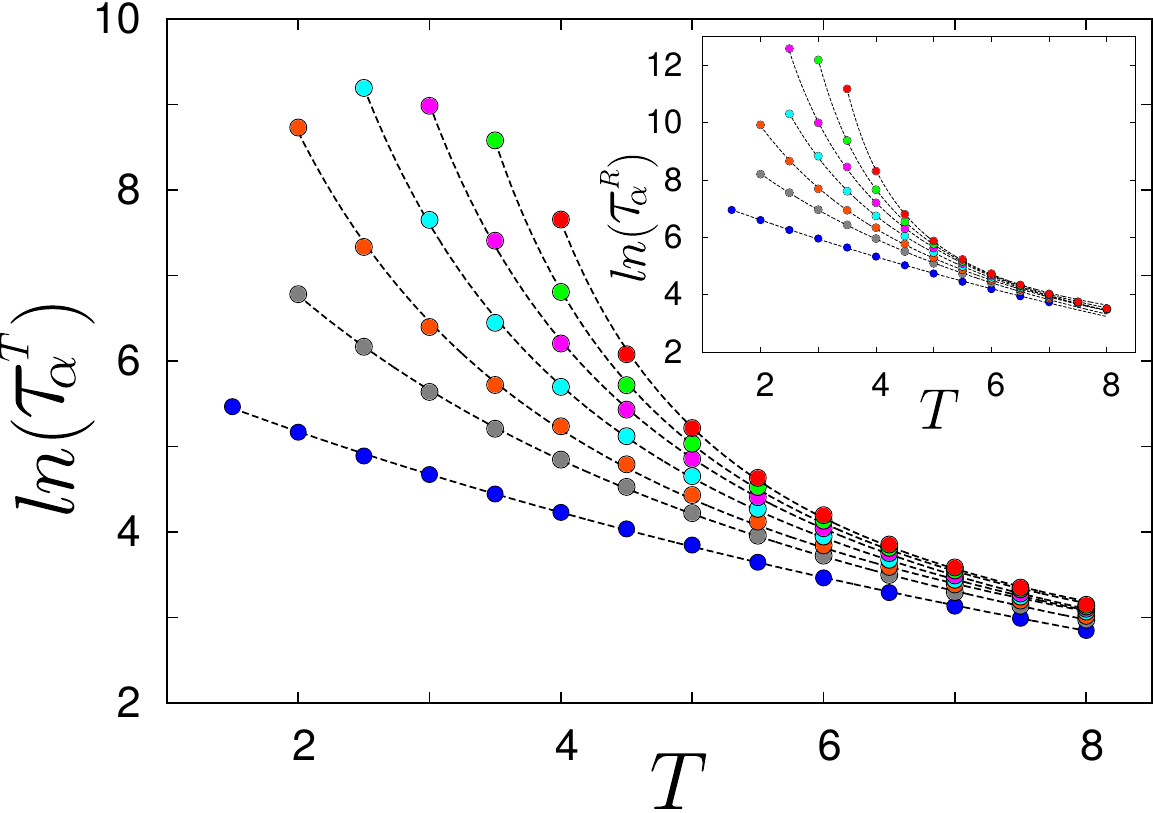}}
\caption{(color online). Translational $\alpha$-relaxation time scale (in filled circles) as a function of temperature ($T$) for different values of self-propulsion force ($f$= $0$ (red), $2$ (green), $3$ (magenta), $4$ (cyan), $5$ (orange), $6$ (dark grey), $8$ (blue)). The black dotted lines represent the fiitting of the data sets for each $f$ to the VFT form. Inset shows rotational $\alpha$-relaxation time scale (in filled circles) as a function of temperature ($T$) for the same values of self-propulsion force ($f$).}
\label{correlator1}
\end{figure}

\begin{figure}
\centerline{\includegraphics[height = 0.4\linewidth,clip=true]{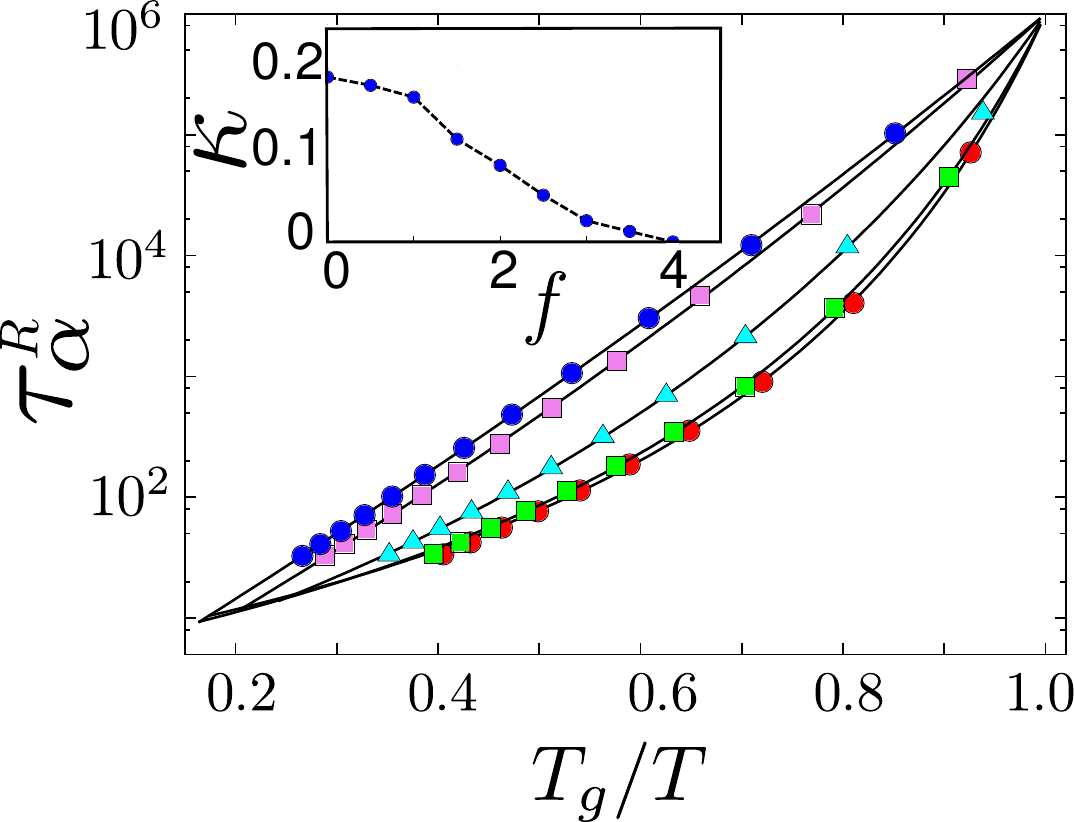}}
\caption{(color online). Variation of  rotational relaxation times $\tau^{R}_{\alpha}$
versus scaled temperature ($T_g/T$) with changing active force : $f=0$ (red), $f=1$ (green), $f=2$ (cyan), $f=3$ (violet), $f=3.5$ (blue). 
$T_g$ is defined as the temperature where $\tau^{R}_{\alpha}=10^6$.
Inset shows how the corresponding kinetic fragility $\kappa$ decreases with increased self-propulsion force.}
\label{rotateangell}
\end{figure}

\begin{figure}
\includegraphics[height = 0.4\linewidth,clip=true]{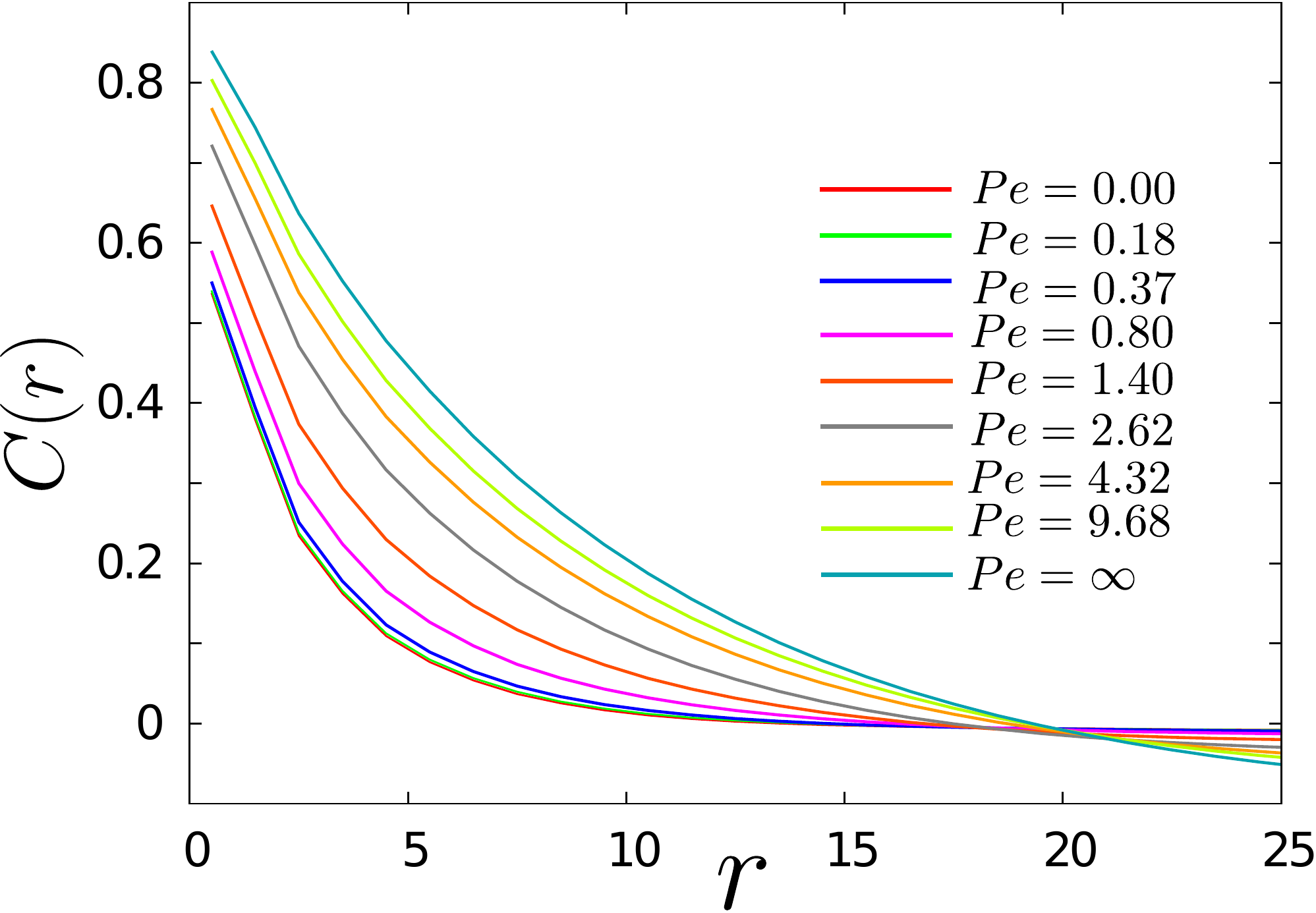}
\caption{(color online). Spatial correlation function of the orientation of the displacement
vectors of the center of mass of the dumbbells,  $C(r)=\frac{1}{2 \pi}\int{\langle 2 \cos^2\Delta \theta( \mathbf{r})-1\rangle} d\phi$, where $\Delta \theta(\mathbf{r})$ is the angular separation between two displacement vectors separated by distance $ \mathbf{r}$. The simulation points are chosen along an iso-$\tau^T_{\alpha}$ line on the ($T-f$) phase diagram. The correlation length $\zeta$ is extracted using the condition $C(\zeta) = 1/e$.}
\label{correlator2}
\end{figure}

\begin{figure}
\includegraphics[height = 0.4\linewidth]{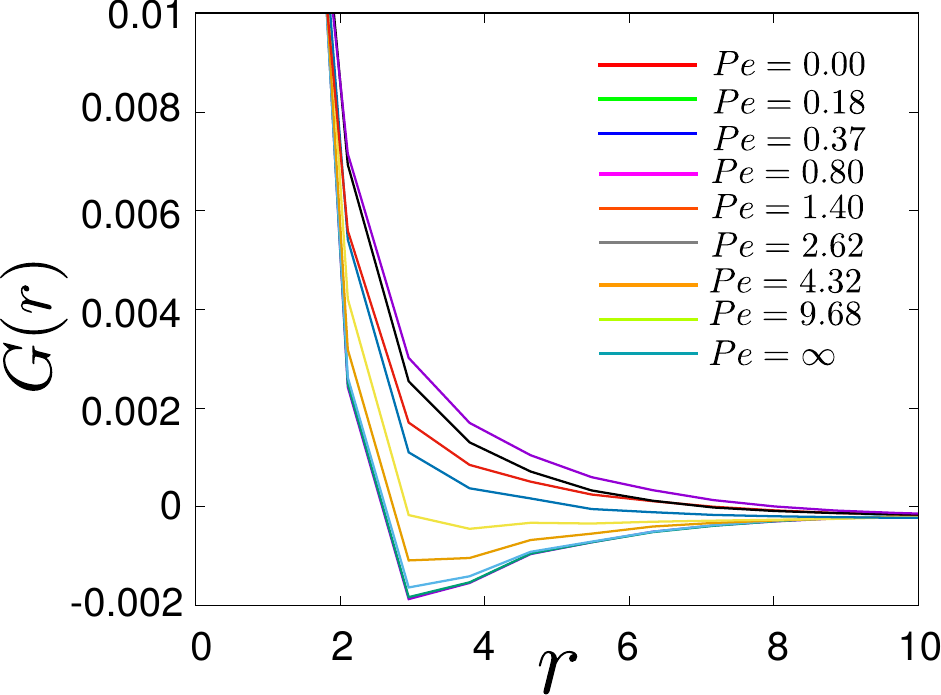}
\caption{(color online). Spatial correlation of the vorticity field of the coarse grained velocity vectors $G(r)=\frac{1}{2 \pi}\int{\langle \omega(\mathbf{0})\omega(\mathbf{r})}  \rangle d\phi$ over a time scale of $\tau^T_\alpha$. The simulation points are chosen along an iso-$\tau^T_{\alpha}$ line on the ($T-f$) phase diagram.
The correlation length $\chi$ is extracted using the condition $G(\chi)=0$.}
\label{correlator3}
\end{figure}

\begin{figure}
\includegraphics[height = 0.4\linewidth]{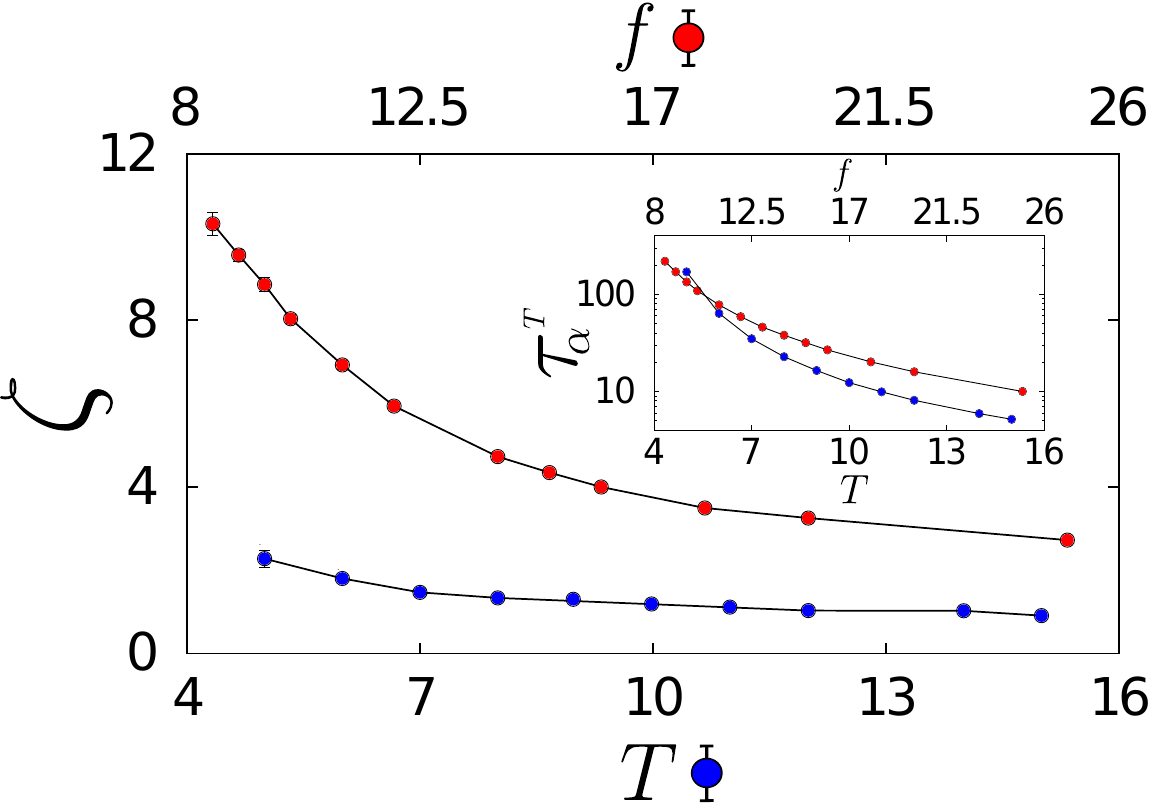}
\caption{(color online). Length scale ($\zeta$) associated with spatial correlation of the direction of the displacement vectors for the dumbbell system at $f=0$ for different $T$ (blue) and at $T=0$ for different $f$ (red), {\textit{i.e.},} when we approach the glass boundary along two different axes. The correlation length scale does not show any significant change if the glass transition is approached by reducing temperature but shows an increase as the system approaches a glass transition by reducing the activity. Inset shows relaxation time scale ($\tau^T_\alpha$) for similar set of simulation points at $f=0$ for different $T$ and at $T=0$ for different $f$.}
\label{lengthscale}
\end{figure}

\begin{figure}
\centerline{\includegraphics[height = 0.4\linewidth]{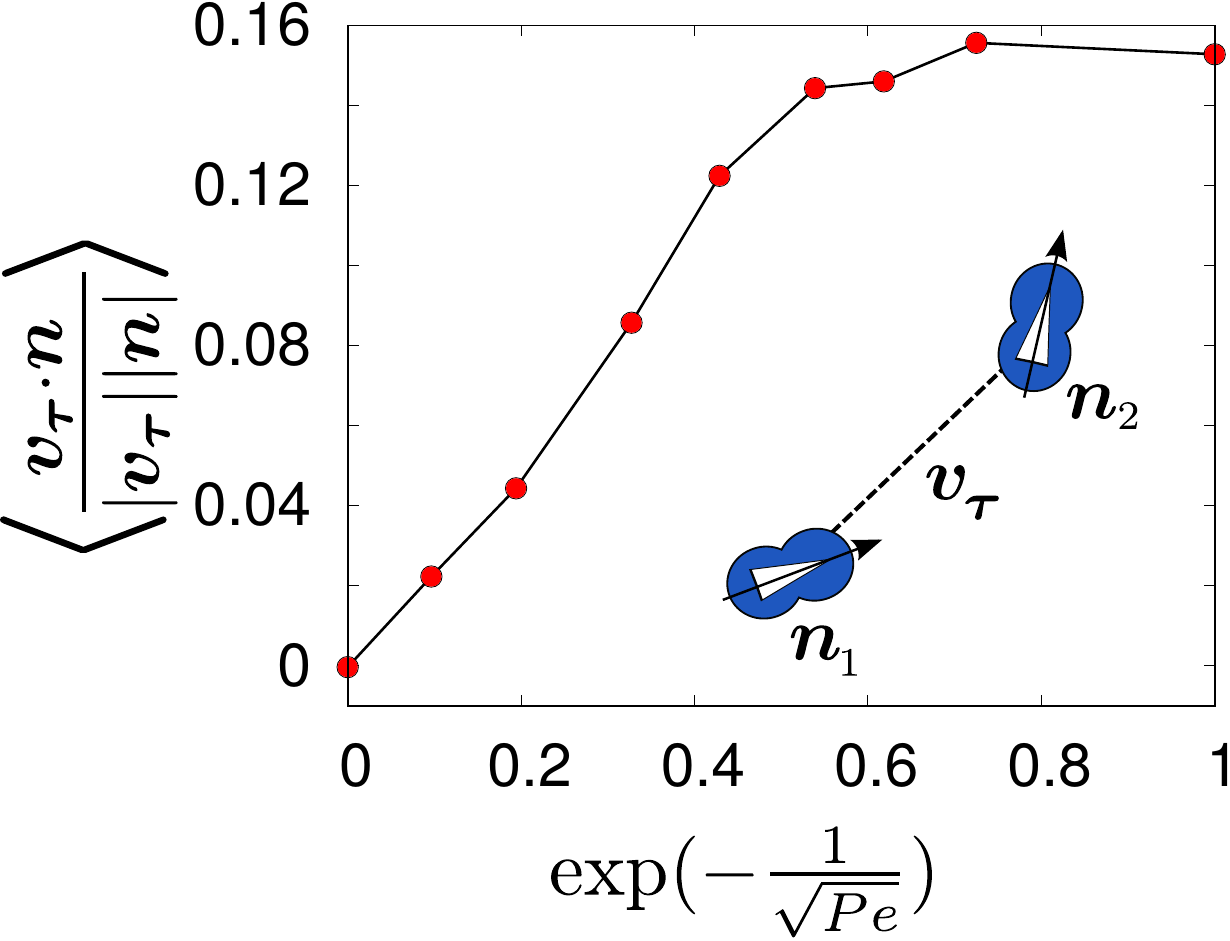}}
\caption{(color online). Average of the normalized dot product of the coarse grained velocity ($\mathbf{v_\tau}$) of each dumbbell and its average orientation vector ($\mathbf{n}=\frac{\mathbf{n1}+\mathbf{n2}}{2}$) over the $\alpha$-relaxation time ($\tau^{T}_\alpha$) as we move through different points on the iso-$\tau^{T}_\alpha$ line from passive to active supercooled liquid region. For few initial points the correlation between displacement and orientation increases and then it saturates as maximum coupling between these two is reached, which is limited by the packing of the dumbbells.}
\label{correlator4}
\end{figure}

\clearpage

\section{Supplementary Movies}

The movies illustrate the time evolution of Voronoi tessellations, constructed from the dumbbell configurations. 
Each Voronoi cell is associated with the center of mass of each dumbbell.
The movies are shown for two points on the iso-$\tau_\alpha$ line
explored in our work, viz. (i) $T=5.4678, f=1$ (weakly active supercooled liquid -- {\it movie1.avi}) and (ii) $T=0, f=10.27$ (strongly active supercooled liquid -- {\it movie2.avi}). For the movies, we focus on
approximately one quadrant of the whole simulation box, in order to properly visualize the topological changes that happen within the system.
In {\it movie1.avi} we mainly observe thermal vibrations with occasional uncorrelated rearrangements of cells, while in {\it movie2.avi} we observe large scale collective rearrangements of cells. As stated in the
main text, this is likely to be relevant for collective cell motion in epithelial tissues.